\documentclass[a4paper,11pt]{article}
\pdfoutput=1
\usepackage{jcappub}
\usepackage{bbold}

\usepackage{amsmath}
\usepackage{linearA}
\usepackage{subfigure}
\usepackage{graphicx}
\usepackage{float}
\usepackage{xcolor}

\def\be{\begin{equation}}
\def\ee{\end{equation}}
\def\bea{\begin{eqnarray}}
\def\eea{\end{eqnarray}}
\def\ba{\begin{align}}

\def\bi{\begin{itemize}}
\def\ei{\end{itemize}}

\def \de {\delta}
\def \ka {\kappa}
\def \Ga {\Gamma}

\def \La {\Lambda}
\def \si {\sigma}

\def \de {\delta}
\def \De {\Delta}

\def \Om {\Omega}
\newcommand{\bx}{{\bf x}}

\newcommand{\bn}{{\bf n}}

\title{Lensing corrections to the $E_g(z)$ statistics from large scale structure}

\author{Azadeh Moradinezhad Dizgah and  Ruth Durrer}
\affiliation{University of Geneva, Department of Theoretical Physics and Center for Astroparticle Physics, 24 quai E. Ansermet, CH-1211 Geneva 4, Switzerland }

\emailAdd{Azadeh.Moradinezhad@unige.ch}
\emailAdd{Ruth.Durrer@unige.ch}

\abstract{We study the impact of the often neglected lensing contribution to  galaxy number counts on the $E_g$ statistics which is used to constrain deviations from GR. This contribution affects both the galaxy-galaxy and the convergence-galaxy spectra, while it is larger for the latter. At higher redshifts probed by upcoming surveys, for instance at $z=1.5$, neglecting this term induces an error of $(25-40)\%$ in the spectra and therefore on the $E_g$ statistics which is constructed from the combination of the two.  Moreover, including it, renders the $E_g$ statistics scale and bias-dependent and hence puts into question its very objective. }

\begin{document}
\maketitle

\section{Introduction}
Understanding the cause of the current accelerated expansion of the universe remains a challenge since its discovery~\cite{Riess:1998cb,Perlmutter:1998np}. Within general relativity (GR), a finely tuned cosmological constant or a dynamical dark energy component can account for the observed acceleration. On the other hand, modifications to GR on cosmological scales can provide an alternative explanation. Distinguishing between these scenarios is one of the main tasks of current and upcoming cosmological large scale structure surveys. At the level of the background, the predictions of most modified gravity and dark energy models are indistinguishable from one another as they can both explain the cosmic acceleration. However in the perturbative regime modified gravity models typically differ from GR. Therefore current and upcoming large scale structure surveys~\cite{Samushia:2013yga,Delubac:2014aqe,DES,SKA,euclid} which probe the evolution of perturbations through weak lensing, redshift space distortions (RSD) and galaxy clustering provide a powerful probe to distinguish between the two ideas. 

Gravitational lensing is sensitive to the sum of the two gravitational potentials $\Phi+\Psi$ while  galaxy clustering of non-relativistic particles is sensitive to the Newtonian potential $\Psi$ only. Therefore the combination of the two observables probes the relation between the two gravitational potentials. Within GR, in the absence of anisotropic stress, which is very small for standard $\La$CDM and quintessence dark energy models, the two are equal. In modified gravity models even in the absence of anisotropic stress, there can be a systematic shift between the two referred to as ``gravitational slip" \footnote{Here we consider modifications of gravity in the Jordan frame. By going to the Einstein frame one can actually move modifications of gravity from a gravitational slip into a fifth force, i.e., a modification of the acceleration equation.}. RSD is in addition sensitive to the rate of growth of structure which in turn depends on the theory of gravity. Combining these three observables into a single statistics, Zhang et al.~\cite{Zhang:2007nk} introduced the $E_g$ statistics which is simultaneously sensitive to modifications to the growth rate, gravitational slip or both. Each of the observables combined in $E_g$ depends on the galaxy bias, i.e., the relation between the galaxies density fluctuations, $\de_g$, and the underlying dark matter  density fluctuations, $\de_m$,  defined by $\de_g=b\de_m$ which in principle is non-linear and scale dependent. On large scales, the biasing relation can be considered to be linear. However, the $E_g$ statistics which is the ratio of galaxy-lensing cross-correlation to the galaxy-velocity cross-correlation is considered to be independent of bias if the same population of galaxies is used to make both measurements. 

There have been several measurements of $E_g$ using various data sets~\cite{Reyes:2010tr,Blake:2015vea,Pullen:2015vtb}. The $E_g$ statistics in these measurements has a slightly different definition than that introduced by Zhang et al. First, the three measurements, used the product of galaxy auto-correlations and $\beta = f/b$ instead of using the galaxy-velocity cross-correlation which is in general difficult to measure. The galaxy-lensing cross-correlation and galaxy auto-correlation are measured from the same survey while $\beta$ is obtained from another survey. Second, in the first two measurements, Fourier-space quantities are replaced by real-space correlation functions. Moreover Pullen et al~\cite{Pullen:2015vtb}, used the CMB lensing-galaxy cross-correlation instead of galaxy-lensing cross-correlations. While the first two measurement of $E_g$ are consistent within $1 \sigma$ with predicted values of GR, the third measurement is in tension with the GR prediction at a level of $2.6 \ \sigma$.

In a recent paper, Leonard et. al.~\cite{Leonard:2015cba} investigated the impact of uncertainties in various theoretical and observational parameters that can modify the GR prediction of $E_g$. In particular they studied how the scale-dependence of galaxy bias will put the efficiency of $E_g$ as a scale-independent measure of deviations from GR into question. In this short paper we discuss an additional difficulty in measuring $E_g$ and we investigate quantitatively its importance for different observational settings.  This is related to the following fact which has recently been put forward~\cite{Yoo:2009au,Yoo:2010ni,Bonvin:2011bg,Challinor:2011bk}: when observing galaxies we cannot directly measure $\de_g(\bx,z)$, but we see photons which have travelled on perturbed geodesics into our telescope. Furthermore, we truly observe the galaxy distribution in angular and redshift space. Whenever we convert this into distances, we make assumptions about the cosmological parameters which we actually want to estimate. For small redshifts, $z\ll 1$ this can be absorbed by measuring distances in units of $h^{-1}$Mpc, where $H_0 =100 h \ {\rm km}s^{-1} {\rm Mpc}^{-1}$ is the present Hubble parameter, but for larger redshifts this relation depends sensitively on all the density parameters, $\Om_m$, $\Om_K$ and $\Om_\La$. As we show here, accounting for these effects, introduces additional contributions to the galaxy-galaxy auto-spectrum and the galaxy-convergence cross-spectrum and in turn renders $E_g$ bias dependent.

The remainder of this paper is structured as follows: In Section~\ref{s:theo} we outline the theoretical background material. In Section~\ref{s:res}  we present numerical results for different observational situations and in Section~\ref{s:con} we conclude.

\section{The theory}\label{s:theo}
We consider only scalar perturbations, for a flat $\Lambda$CDM universe described by a perturbed Friedmann-Robertson-Walker metric. We use the following notation for the metric in the conformal Newtonian gauge
\be
ds^2 = a^2(t)\left[-(1+2 \Psi) dt^2 + (1-2\Phi) \delta_{ij} dx_i dx_j\right]\,.
\ee 
Within GR, in the absence of anisotropic stress, the two gravitational potentials are equal $\Phi=\Psi$. Moreover the gravitational potential $\Phi$ is related to the matter over density, $\delta_m$ in comoving gauge, via the Poisson equation
\be
k^2\Phi = 4\pi G \bar\rho a^2\de_m(k,z) =\frac{3}{2}H_0^2(1+z)\Om_{0,m}\de_m(k,z)\,.
\ee
Modified gravity models generally change this relation in addition to changing the relation between the gravitational potential and the matter over density via changing the Poisson equation. Since lensing is sensitive to the projected matter density, more precisely  to $\nabla^2 (\Phi+\Psi)$ while the peculiar velocity field of non-relativistic particles is sensitive to the time-component of the metric and hence $\Psi$, the ratio of the two can be used to test the relation between the two gravitational potentials. This is the basis for the $E_g$ statistics defined as~\cite{Zhang:2007nk} 
\be \label{eq:E_g_def}
E_g(z,k) = \frac{k^2 (\Phi+\Psi)}{3 H_0^2(1+z)\theta(k)}.
\ee
It is sensitive to deviations from GR through its sensitivity to the relation between the two gravitational potentials and the growth rate. Here $\theta$ is the divergence of the peculiar velocity field which on linear scales, in a $\La$CDM  cosmology is given by $\theta(k) = f(z) \delta_m(k,z)$ where $f(z)$ is the growth rate. Within GR it is given by $f(z) \simeq [\Omega_m(z)]^{0.55}$. Assuming $\La$CDM cosmology, $E_g$ simply becomes $E_g(z,k) = \Omega_{m,0}/ f(z)$ which on linear scales is therefore a scale-independent quantity. At linear level, modification to GR can be phenomenologically parameterized by introducing two arbitrary functions $\mu(k,z)$ and $\gamma(k,z)$ such that
\bea
 k^2\Phi &=& 4\pi G \bar\rho a^2 \mu(k,z) \de_m(k,z)  \,,\nonumber \\
 \Psi &=& \gamma(k,z) \Phi   \,.
\eea
$E_g$ is related to these parameters of modified gravity models via
\be
E_g(k,z) = \frac{\Omega_{m,0} \ \mu(k,z)\left[\gamma(k,z)+1\right]}{2f} \,.
\ee
Note that the growth rate is typically also modified with respect to that of GR. Modifications of GR therefore in general render $E_g$ scale dependent or at least change its redshift dependence. This, and its bias-independence make it a useful statistics to test for modifications of GR.

Given the definition of $E_g$, one can construct various estimators to extract the information about the gravitational potentials. Zhang et. al.~\cite{Zhang:2007nk}  proposed using cross-correlations of galaxies with the shear of the background galaxies, i.e., using the galaxy-galaxy lensing and the cross correlations of galaxies with the velocity field using RSD. Reyes et al.~\cite{Reyes:2010tr} implemented this consistency test by considering the product of the galaxy-galaxy auto-power spectrum and the RSD parameter $\beta = f/b_g$  obtained from independent measurements instead of the galaxy-velocity cross-spectrum measured from RSD. Recently Pullen et al.~\cite{,Pullen:2015vtb} advocated using the CMB convergence-galaxy cross-correlation instead of galaxy-galaxy lensing as a more robust lensing tracer which can probe $E_g$ at higher redshifts and larger scales than what is possible with galaxy lensing. For the rest of this paper we discuss this estimator.

Due to lensing by the intervening matter, the CMB photons observed in direction ${\bf n}$ were emitted in directions ${\bf n} + \nabla \phi( {\bf n})$, where $\phi( {\bf n})$ is the lensing potential. The lensing convergence $\kappa(\bn,z)$ is the determinant  of the lens map and in multipole space,  related to the lensing potential by $2 \kappa_{\ell m} = \ell(\ell+1)\phi_{\ell m} $, it is given by
\be
\kappa(\bn,z)= \frac{1}{2} \int_0^{\chi(z)} d\chi \ \frac{\chi(z)-\chi}{\chi(z)\chi} \ \nabla^2_\Omega(\Phi+\Psi)(\chi(z)\bn,t) =   \frac{1}{2}\nabla^2_\Omega\phi(\bn,z) \,,
\ee
where $\chi(z)$ is the comoving distance to redshift $z$. The conformal time is simply $t=t_0-\chi(z)$, where $t_0$ is present time. The operator $\nabla^2_\Omega$ indicates the Laplacian on the sphere. For the CMB lensing signal $z=1080\equiv z_*$. The convergence-galaxy cross-spectrum is then given by
\be
\langle \kappa(\bn,z_*)\de_g(\bn',z')\rangle = \frac{1}{2}\int_0^{\chi(z_*)} d\chi \ \frac{\chi(z_*)-\chi}{\chi(z_*)\chi}\ \langle\nabla^2_\Omega(\Phi+\Psi)(\bn\chi,t_0-\chi)\delta_g(\chi(z')\bn',t_0-\chi(z'))\rangle \,.
\ee
Expanding the convergence and the galaxy fields into spherical harmonics
\bea
\kappa(\bn,z) &=& \sum_{\ell m}a_{\ell m}^\ka(z) Y_{\ell m}(\bn), \nonumber \\
\de_g(\bn,z) &=& \sum_{\ell m}a_{\ell m}^{g}(z) Y_{\ell m}(\bn), 
\eea
the galaxy-convergence angular cross-spectrum is defined in terms of the cross-correlation as
\be
\langle  \kappa(\bn,z) \de_g(\bn',z')\rangle = \frac{1}{4\pi}\sum_\ell (2\ell+1)C_\ell^{\ka g}(z,z')P_\ell(\bn\cdot\bn')\,.
\ee
where  $P_\ell$ is the Legendre polynomial of degree $\ell$ and the convergence-galaxy cross-spectrum is given by
\be
C_\ell^{\kappa g}(z,z') = \langle a_{\ell m}^\ka(z)a_{\ell m}^{g *}(z')\rangle,\
\ee
Correspondingly one can define the galaxy-galaxy power spectrum
\be
C^{g g}_\ell(z,z') = \langle a_{\ell m}^g(z)a_{\ell m}^{g *}(z')\rangle\,.
\ee
Using these two observables in addition to the measurement of the parameter $\beta$ from the redshift space distortions in the power spectrum, we can construct the the $E_g$ estimator as ~\cite{Pullen:2015vtb} 
\be 
\label{eq:E_g_es}
E_g(z,\ell) = \Gamma \frac{C_\ell^{\kappa g}(z_*,z)}{\beta C_\ell^{gg}(z,z)}.
\ee
$\Ga$ is a factor correcting for the window function of the lensing integral and the redshift distribution of the observed galaxies (see~\cite{Pullen:2015vtb} for details).  Both, $C_\ell^{\kappa g}$ and $\beta C_\ell^{gg}$ are linear in the galaxy bias $b$ so that  the ratio $E_g$ is bias independent. At present, the dominant uncertainty in this measurement comes from  CMB lensing noise. Using the cross-correlation of the Planck CMB lensing map with the Sloan Digital Sky Survey III (SDSS-III), the CMASS galaxy sample, and combining this with measurements of the CMASS auto-power spectrum and the RSD parameter $\beta$, Pullen et al.~\cite{Pullen:2015vtb} find $E_g(z=0.57) = 0.243 \pm 0.060 (stat) \pm 0.013 (sys)$  which is in tension with the GR prediction $E_g(z = 0.57|{\rm GR}) = 0.402 \pm 0.012$ by $2.6 \sigma$. 

In the present paper we show, that the measurements performed for $E_g(z)$ do not exactly measure what has been calculated. They are affected by additional lensing terms appearing whenever we observe galaxies at appreciable redshifts. This observation which we detail and quantify below also affects other proposals to measure $E_g(z)$ which have been presented in~\cite{Reyes:2010tr,Blake:2015vea,Leonard:2015cba}.

In Refs.~\cite{Bonvin:2011bg,Challinor:2011bk} it has been shown that when we correlate number density fluctuations of galaxies at `high redshift', i.e. $z\gtrsim 0.5$ or at very large scales, relativistic effects coming from volume perturbations and from the fact that observations are made on the perturbed background lightcone have to be taken into account. The simplest of these are the well known redshift space distortion~\cite{Kaiser:1987qv} which represents the longitudinal deformation of the volume and lensing~\cite{Matsubara:2000pr} which is the transversal volume deformation including a possible magnification bias. If relatively large redshift bins are considered, $\De z\gtrsim 0.1$, redshift space distortion can be completely neglected and if all the scales involved are significantly subhorizon, say $\ell>20$ also the effects involving the gravitational potential directly (integrated Sachs-Wolfe effect, Shapiro time delay etc.) can be ignored, however, the term coming from lensing has to be included so that the galaxy fluctuations in direction $\bn$ and at redshift $z$ can be approximated by (see for instance \cite{DiDio:2013bqa} for the full expression of $\Delta^g$ at linear order)
\be\label{eq:Deltag}
\Delta^g(\bn,z) =  b\de_m(\chi(z)\bn,z) + \left(1-\frac{5}{2}s\right)\int_0^{\chi(z)}d\chi \ \frac{\chi(z)-\chi}{\chi(z)\chi}\ \nabla^2_\Om(\Phi+\Psi)(\chi\bn,t_0-\chi)\,.
\ee
 $b$ is the galaxy bias relating the galaxy overdensity to the underlying dark matter. For a flux-limited survey, $s$ is the magnification bias describing the slope of the luminosity function
\be
 s(z,m_{\rm lim}) \equiv \left.-\frac{2}{5}\frac{{\rm d ln} \ \bar n_g(z,L>L* )}{{\rm d ln \ L}}\right|_{L*}
\ee
where $L*$ is the luminosity threshold. 

For $\La$CDM, setting $\Phi=\Psi$ and using Limber approximation~\cite{LoVerde:2008re} one can write the angular power spectra in terms of the primordial power spectrum $P(k)$ as

\bea \label{eq:galaxy_conv}
C_\ell^{gg}(z,z) &=& \frac{1}{2\pi^2}\int_0^\infty dk \ k^2 P(k) W_\ell^g(k,z)W_\ell^g(k,z) \\
C_\ell^{\kappa g}(z_*,z) &=& \frac{1}{2\pi^2} \int_0^\infty dk \ k^2 P(k) W_\ell^\kappa(k,z_*)W_\ell^g(k,z) \,.
\eea
The kernels of the galaxy number count and the CMB lensing are given by
\bea
W_\ell^\kappa(k,z_*) &=& \frac{3\Omega_{m0} H_0^2}{2} \int_0^{z_*} \frac{dz}{H(z)}\  \frac{\chi(\chi_*-\chi)}{\chi_*}(z)D(z)j_\ell(k\chi(z)) \\
W_\ell^g(k,z) &=& \int^{z_*}_0 dz \ w(z) \frac{dN}{dz}  T^g_\ell(z,k) \label{e:Wgnaif}
\eea
 where $w(z)$ is the window function describing the redshift bin in a given survey and $\chi_*=\chi(z_*)$.  $D(z)$ is the growth function defined by
 $$
 \de_m(z,k) = \frac{D(z)}{1+z}\de_0(k) \,,
 $$
 where $\delta_0(k)$ is the (linear) density fluctuation today and $P(k)$ is its power spectrum. 
 $dN/dz$ is the redshift distribution of the galaxies considered. The transfer function $T^g_\ell(z,k)$ is given by
 $$
T^g_\ell(z,k)=  \left[j_\ell(k\chi(z)) b\frac{D(z)}{1+z} +(2-5s)W_\ell^\ka(k,z)\right]\,,
 $$
 where the galaxy bias, $b$  in general depends on redshift and on scale. This includes the terms of Eq.~(\ref{eq:Deltag}), assuming $\Phi=\Psi$. It is a good approximation when the redshift slice is rather wide so that  redshift space distortions can be neglected and for $\ell>20$ so that also contributions from the gravitational potential can be neglected. In our numerical results we show also the other relativistic contributions and the contribution from RSD. The full formula for all terms is given in~\cite{DiDio:2013bqa} and we do not repeat it here.

The second term in $T^g_\ell(z,k)$ comes from the fact that the transversal volume (or better the area) seen under a given opening angle is modified by lensing of foreground galaxies~\cite{Bonvin:2011bg}. If not all galaxies are seen but some `make it into the survey' due to magnification by lensing, there is an additional contribution from magnification termed `magnification bias'~\cite{Challinor:2011bk}. We neglect this in the present discussion, but depending on the details of a given survey this term can be very relevant, see e.g.~\cite{Montanari:2015rga}.

Clearly, the second term of (\ref{eq:Deltag}) is neglected in the analysis of~\cite{Pullen:2015vtb}. This term also contributes to the galaxy-galaxy lensing, i.e. the cross-correlation of the shear of a background galaxy with the lens galaxy, $C_\ell^{\ka g}$ in the same way and thus affects the other estimator used by Reyes et al. . For brevity and simplicity we concentrate here on the analysis presented in~\cite{Pullen:2015vtb}. Note that in the correct analysis the $C_\ell^{\kappa g}(z_*,z)$ term contains also a $2C_\ell^{\kappa \ka}(z_*,z)$, while the $C_\ell^{gg}(z,z)$ term contains in addition to the expression in eq.~(\ref{e:Wgnaif}) $4C_\ell^{\ka g}(z,z) +4C_\ell^{\ka\ka}(z,z)$.  Here $z$ denotes the redshift of the galaxy survey.
 
 For what follows, we assume a Gaussian window with half width  $\Delta z = 0.1$.  Moreover we consider two different galaxy redshift distributions: for low redshifts, we consider a distribution for a DES-like survey, 
\be \label{eq:DES_dNdz}
\frac{d N}{dz} \propto \left(\frac{z}{z_g}\right)^2 \exp\left[-\left(\frac{z}{z_g}\right)^2\right]
\ee
and for high redshifts, we utilize a normal distribution \cite{Bianchini:2015fiw, Bianchini:2015yly},
\be \label{eq:half-normal}
\frac{d N}{dz} = \frac{1}{\sigma \sqrt{2\pi}} {\rm exp}\left(-\frac{(z-z_g)^2}{2\sigma^2}\right) \,.
\ee
The results we present in the next section, only weakly depend on the assumed galaxy distribution. For instance at z = 1.5 we have also computed the same quantities for the DES-like distribution defined in Eq. (2.19) with very similar results.

\section{Results}\label{s:res}
In this section we show that the most relevant effect is neglecting $2C_\ell^{\kappa \ka}(z_*,z)$ in the expression for $C_\ell^{\ka g}$.  Even though it accounts for less than 5\% in a measurement with $z_g=0.57$, it goes in the right direction since  $\ka$ and $\de$ are anti-correlated so that this term reduces the signal. We shall also see that the contribution becomes more important at higher redshift and therefore certainly cannot be neglected in future surveys.

We work within linear perturbation theory and compute all the terms missing in the standard analysis. In addition to the lensing contribution, we also compute the contribution from redshift space distortions and the gravitational potential terms to ensure a control over our approximations. We use the CLASS code~\cite{Blas:2011rf,DiDio:2013bqa} to calculate the spectra. We assume a six-parameter cosmology with $n_s=  0.9682, A_s = 2.2006 \times 10^{-9},~ \tau = 0.079,~  \Omega_bh^2 = 0.02225,~  \Omega_ch^2 = 0.1194 \  {\rm and} \  h = 0.6748$. For simplicity,  we set the galaxy bias to $b=1$ and magnification bias to $s=0$. It should be noted that the exact value of the galaxy bias and magnification bias, determines the relative importance of the lensing and density contributions. Therefore one has to input the values for these parameters for the corresponding redshift and galaxy population of a given survey.

The relative errors on $C_\ell^{\kappa g}$ and $C_\ell^{gg}$ are defined as
\be
\frac{\Delta C_\ell^{\kappa g}}{C_\ell^{\kappa g}} = \frac{C_\ell^{\rm lens-dens} - C_\ell^{\rm lens-all}}{C_\ell^{\rm lens-dens}}, \qquad \qquad \frac{\Delta C_\ell^{g g}}{C_\ell^{g g}} = \frac{C_\ell^{\rm dens-dens} - C_\ell^{\rm all-all}}{C_\ell^{\rm dens-dens}},
\ee
where $C_\ell^{\rm lens-dens}$ denotes the correlation of density with $\kappa$ while $C_\ell^{\rm lens-all}$ is the cross-correlation of the full galaxy number counts at first order (see for instance \cite{DiDio:2013bqa}) with $\kappa$. $C_\ell^{\rm dens-dens}$ refers to the galaxy auto-correlation when only the density contribution is accounted for while $C_\ell^{\rm all-all}$ is calculated using the full first-order expression for the galaxy number counts.

\begin{figure}\vspace{-.1in}
\hspace{-.6in}\includegraphics[width=1.2\textwidth,height=4.6in]{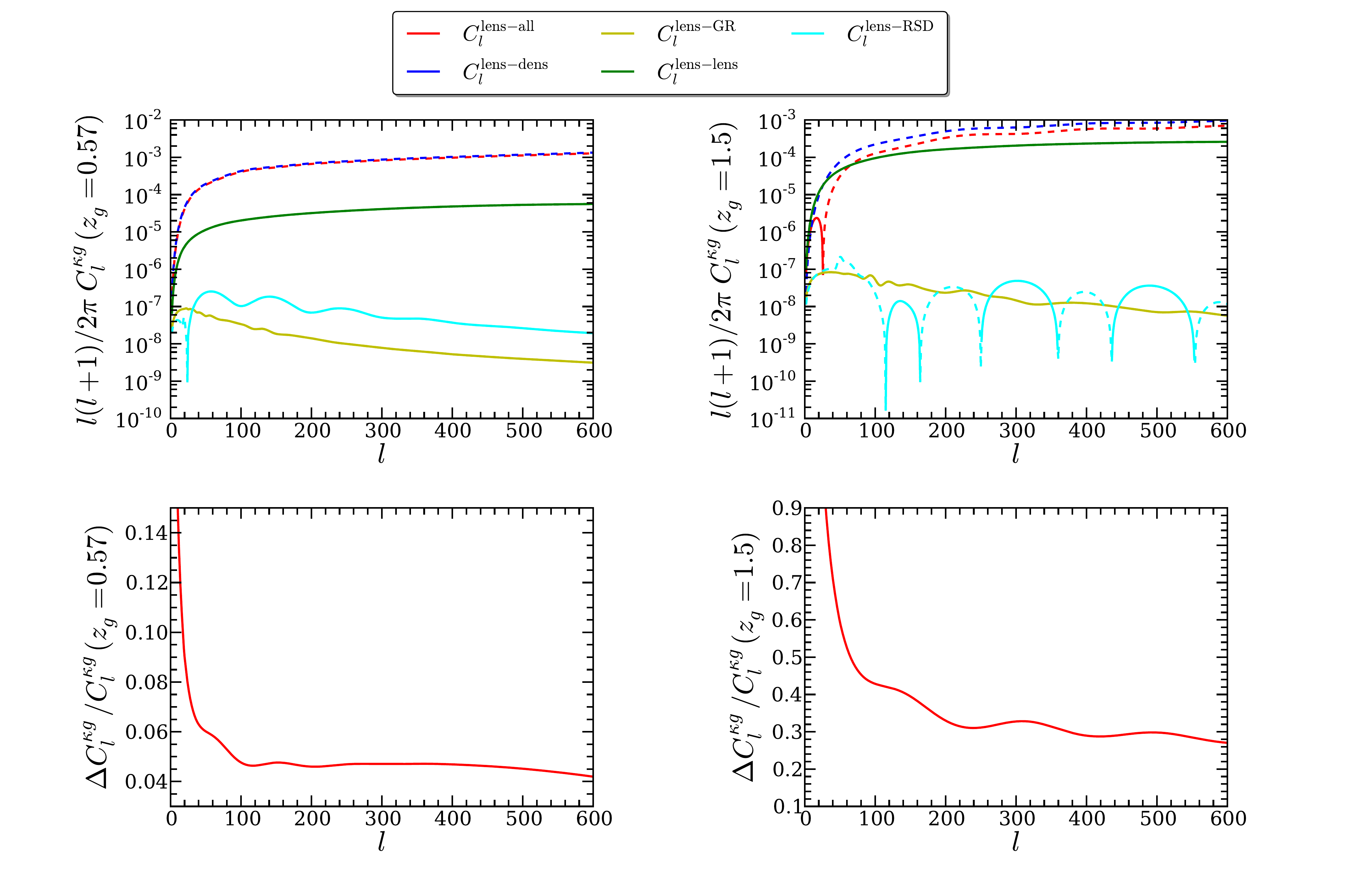}
\vspace{-.3in}
\caption{Top panel shows the convergence-galaxy cross-spectrum when the galaxies are taken at $z_g=0.57$ on the left and $z_g=1.5$ on the right panel. Different lines correspond to different contributions to the galaxy number counts, dashed lines denote negative correlations. The bottom panel shows the relative error induced if only the density contribution to galaxy number count is accounted for.  For the matter power spectrum, the non-linear spectrum obtained from halo-fit is used. The redshift distribution of galaxies $dN/dz$ at $z_g=0.57$ is taken to be that of a DES-like survey, (\ref{eq:DES_dNdz}),  while for $z_g=1.5$ a normal distribution, (\ref{eq:half-normal}), with $\sigma=0.68$ is used. These cross-spectra are calculated for $b=1$ and $s=0$. The latest measurement error from DES science-verification data   \cite{Chang:2016npo} and H-ATLAS galaxies \cite{Bianchini:2015yly} is of the order of $10\%$ which is expected to improve.} 
\label{fig:kappag}
\end{figure}

In Fig.~\ref{fig:kappag} we plot different contribution to $C^{\ka g}_\ell$ for galaxy sample at two different redshifts $z_g=0.57$ and $z_g= 1.5$. Clearly, the dominant term is the correlation of the density with $\ka$ (denoted `lens-dens') which is taken into account in the standard analysis. But the `lens-lens' term, $2C^{\ka \ka}_\ell$ amounts to nearly 5\% already at $z=0.57$ and becomes up to $(25$--$40)$\%  at $z=1.5$. Note that for $z=1.5$ at $\ell \lesssim 25$ the  `lens-lens' term even dominates and changes the sign of the total signal. 

Therefore, even though this contribution cannot explain the $2.6\si$ discrepancy observed by~\cite{Pullen:2015vtb}, it goes in the right direction and reduces the tension to below $2\si$.  What is more important is that in order to take advantage of the wealth of information from future surveys such as Euclid, which probe higher redshifts, to constrain $E_g$ with higher precision, we definitely have to include the lensing term. Unfortunately with this contribution the ratio $C^{\ka g}/(\beta C^{gg})$ is no longer bias-independent and so this quantity looses its interest to a large extent. The contribution from redshift space distortions (denoted `lens-RSD') and the one from the gravitational potential terms (denoted `lens-GR') remain very small on all scales.

\begin{figure}\vspace{-.045in}
\hspace{-.6in}\includegraphics[width=1.2\textwidth,height=4.in]{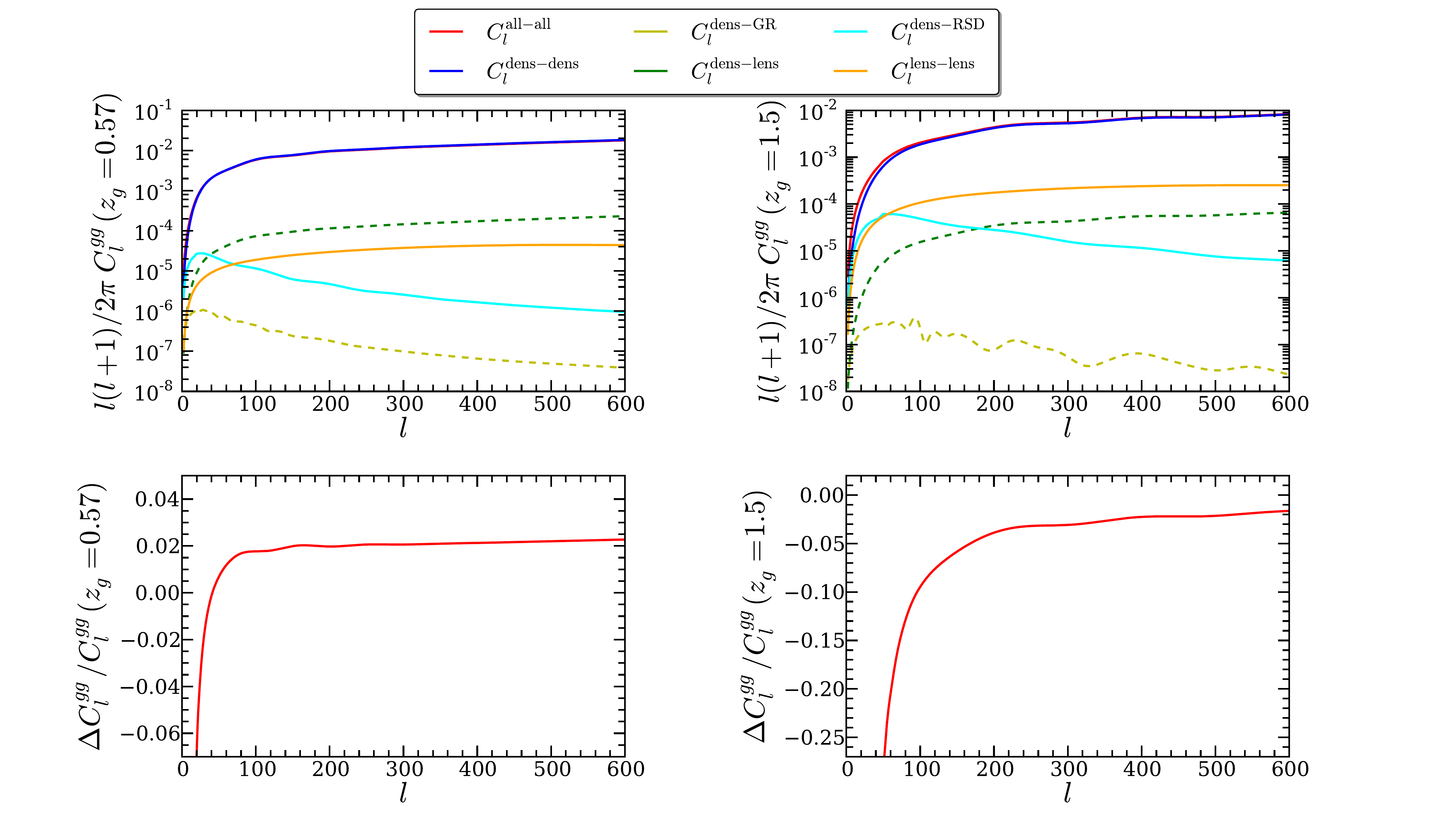}
\vspace{-.3in}\caption{The galaxy-galaxy auto-correlation (top panel) and the relative error (bottom panels) are shown. The galaxies are at the same redshifts $z_g=0.57$ on the left and $z_g=1.5$ on the right. Here 'all-all' is the galaxy auto-correlation when all the contributions are included, while 'dens-X'  (where X can be dens, lens, GR and RSD) refer to the correlation between the density term with each of the other contributions, and 'lens-lens' refers to the correlation between the lensing contributions in both terms.  The non-linear matter density employing halo-fit is used.  The redshift distributions of galaxies $dN/dz$ are the same as in Fig~\ref{fig:kappag}.  These cross-spectra are calculated for $b=1$ and $s=0$. The latest measurement error from DES science-verification data   \cite{Chang:2016npo} and H-ATLAS galaxies \cite{Bianchini:2015yly} is of the order of $10\%$ which is expected to improve.}
\label{fig:gg}
\vspace{-.1in}
\end{figure}

Neglecting the lensing terms also affects the galaxy-galaxy auto-spectrum. The effect is however smaller than that on convergence-galaxy cross-spectrum. In Fig.~\ref{fig:gg} we show the contribution from different terms to the signal. The effect of the new terms at $\ell>100$ is on the level of 2\% for  $z_g=0.57$ and only slightly larger for $z_g=1.5$. Notice that the sign of the relative error changes, i.e. at lower redshift, including the other contributions in addition to the `density-density' term, reduces the galaxy-galaxy spectrum while for the higher redshift, it is increased. This is due to the fact that while at lower redshift, the RSD contribution and also the lens-lens term which are both positive are negligible for $\ell>50$, while at higher redshift these contributions, especially the lens-lens term become more important and more than compensate the negative dens-lens term so that now there is an increase in the galaxy-galaxy spectrum with respect to  only accounting for the `dens-dens' term. 

\begin{figure}\label{fig:Eg_ratio}
\hspace{-.3in}\includegraphics[width=1.1\textwidth,height=2.in]{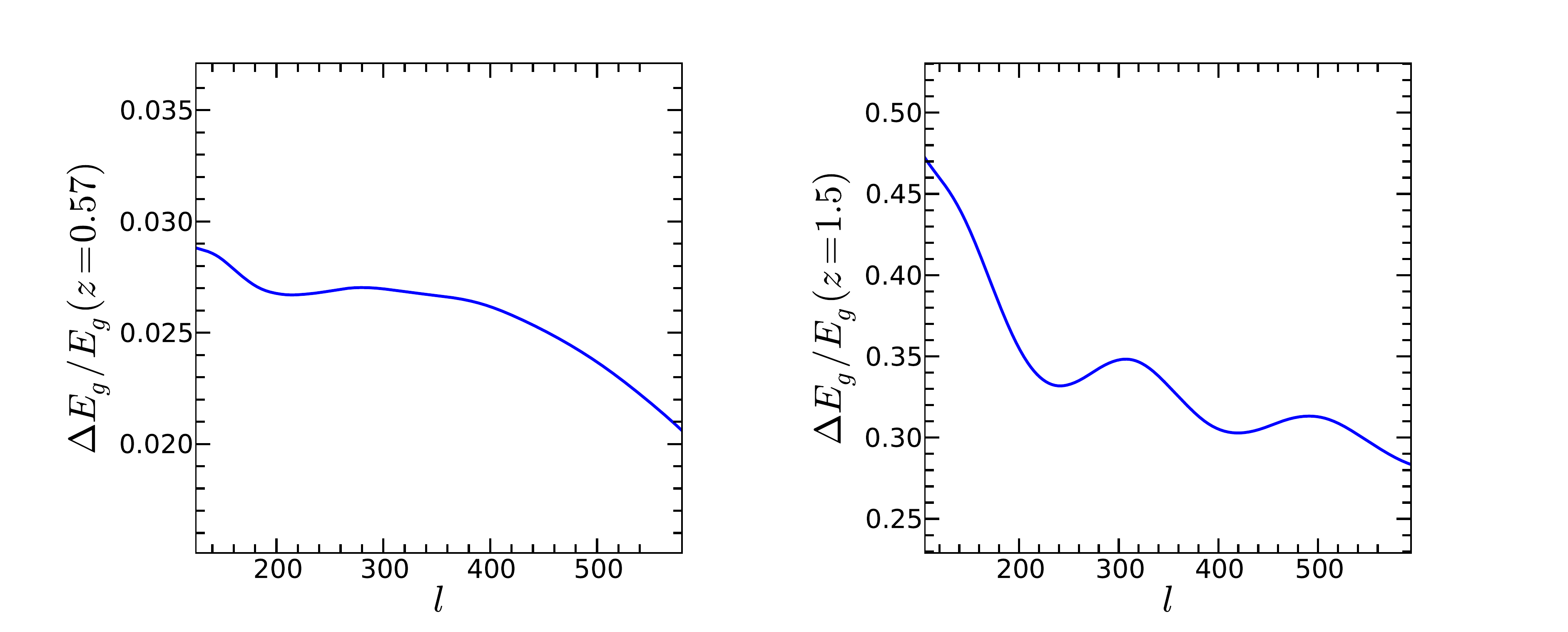}
\caption{The relative error of the $E_g$ for $z=0.57$ on the left and $z=1.5$ on the right. The choice of $dN/dz$ is the same as in the previous plots. The error is calculated assuming $b=1$ and $s=0$.}
\end{figure}

In Fig.~\ref{fig:Eg_ratio} we plot the relative error on $E_g$ defined as 
\be
\frac{\Delta E_g}{E_g} = 1- \frac{C_\ell^{\rm dens-dens}}{C_\ell^{\rm lens-dens}} \frac{C_\ell^{\rm lens-all}}{C_\ell^{\rm all-all}}.
\ee
Including the additional terms in the convergence-galaxy and galaxy-galaxy cross and auto spectra, renders the $E_g$ scale and bias dependent and induces an error of about $(2-3) \%$ at $z=0.57$ and of about $(30-45)\%$ at $z=1.5$. At $z = 0.57$, the error induced by neglecting these additional terms is well within the statistical error of the measurement of Pullen et al. which is about $25\%$. It can also be accounted for as an additional systematic error to their currently reported value which is about $5\%$. At higher redshifts probed by the upcoming surveys which are expected to obtain higher precision measurements, neglecting this term will bias the theoretical estimate of $E_g$. 

\begin{figure}[H]\label{fig:rel_err_bs}\
\centering
\includegraphics[width=0.8\textwidth]{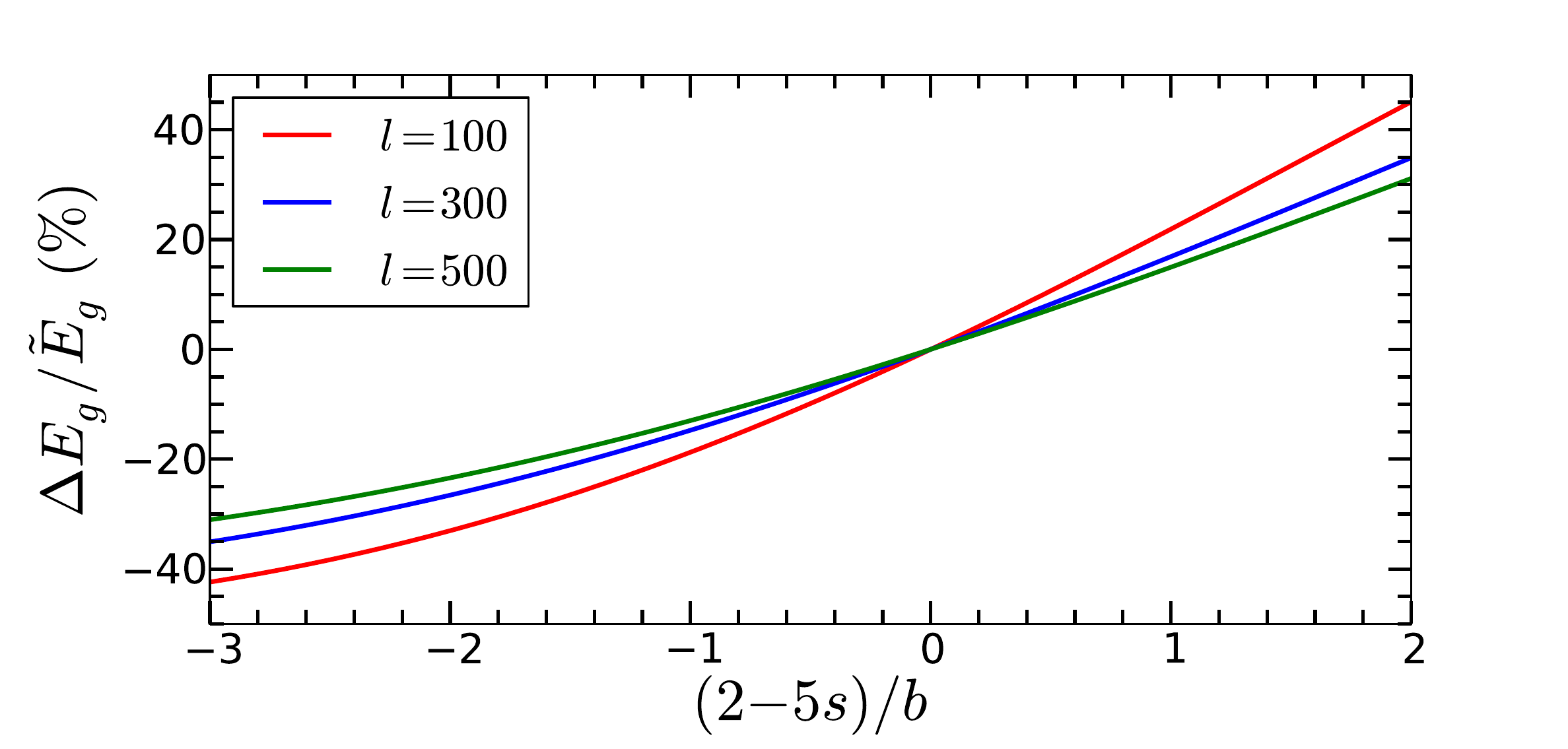}
\caption{Relative error induced on $E_g$ due to neglecting the lensing contribution at $z=1.5$ as a function of $(2-5s)/b$. The choice of $dN/dz$ is the same as in the previous plots.}
\end{figure}

As we noted in the beginning of this section, the exact value of the corrections to the $E_g$ due to these additional effects, largest of which is lensing contribution to the galaxy number counts, depends on the value of galaxy bias and magnification bias. For instance accounting only for the additional lensing term we have
\begin{eqnarray}
E_g \propto \frac{C_\ell^{\kappa g}}{\beta C_\ell^{gg}} &=& \frac{b C_\ell^{\kappa \delta}+(2-5s)C_\ell^{\kappa \kappa_g}}{b^2 C_\ell^{\delta \delta}+(2-5s)^2C_\ell^{\kappa_g \kappa_g}+2b(2-5s)C_\ell^{\kappa_g \delta}} \nonumber\\ 
\nonumber \\
&=& \tilde E_g \ \frac{1+\left(\frac{2-5s}{b}\right) r_1}{1+\left(\frac{2-5s}{b}\right)^2 r_2 + 2\left(\frac{2-5s}{b}\right)r_3},   \label{eq:sb_dep}
\end{eqnarray}
where
\begin{equation}
r_1 = \frac{C_\ell^{\kappa \kappa_g}}{C_\ell^{\kappa \delta}},  \qquad \qquad r_2 = \frac{C_\ell^{\kappa_g \kappa_g}}{C_\ell^{\delta \delta}}, \qquad \qquad r_3 = \frac{C_\ell^{ \kappa_g \delta}}{C_\ell^{\delta \delta}}, \qquad \qquad \tilde E_g \propto \frac{C_\ell^{\kappa \delta}}{\beta b C_\ell^{\delta \delta}}.
\end{equation} 
In the above expression $\kappa$ denotes the CMB lensing while $\kappa_g$ denotes the lensing of background galaxies by the foreground galaxies. In the often encountered case where $r_1\gg r_2,r_3$, the correction is simply dressed by a factor $(2-5s)/b$ which depends on the survey specifications. To highlight the dependance of the error on $b$ and $s$, in Fig.~\ref{fig:rel_err_bs} we plot the relative error defined as
\be\label{eq:err-def}
\frac{\Delta E_g}{\tilde E_g} = 1-\frac{E_g}{\tilde E_g}
\ee
as a function of $(2-5s)/b$ at $z=1.5$. Different lines correspond to different values of $\ell$. As is also clear from eqs.~(\ref{eq:sb_dep}) and (\ref{eq:err-def}), for a galaxy population with $(2s-5)/b=0$ the error is reduced to zero.  It is therefore best to consider surveys with $s\sim 0.4$ and large bias for this test.

\section{Conclusions}\label{s:con}
The $E_g$ statistics which combines the information from lensing, galaxy clustering and RSD observables into a single statistics, has been proposed and used as a promising measure to constrain deviations from GR. As the precision of the measurements is increasing, the importance of some of the effects that have been conventionally neglected in theoretical predictions of $E_g$ must be reconsidered. In this short paper, we point out  one such effect, which is the contribution of lensing to the galaxy-clustering signal which has not been accounted for in previous works.

In galaxy surveys, the number of galaxies at a given redshift and angle on the sky is obtained by counting the galaxies within the observed volume. Since we observe galaxies via the photons that have traveled to us on our background lightcone, the matter distribution in between, deviates the photon trajectory and hence the angle at which a given galaxy is observed. Moreover, the redshift at which galaxies are observed is is also modified. The change of the angle and redshift induces various physical effects that modify the observed galaxy distribution by introducing additional contributions to the well-known density perturbations and redshift-space distortion. For instance there will be also a lensing term and several other contributions from the gravitational potentials. 

Here, we study the impact of neglecting these contributions, the dominant of which is the lensing. We show that neglecting the lensing of number counts induces a theoretical error for the prediction of both the convergence-galaxy cross-spectrum and the galaxy-galaxy auto-spectrum. We consider the CMB convergence as the lensing tracer field. Our results indicate that the size of the error depends on the redshift of the galaxy sample. At $z=0.57$, the relative error induced at angular scales of $\ell>100$ is of less than $5\%$ for convergence-lensing cross-spectrum and about a factor of 2 smaller for the galaxy-galaxy auto-spectrum. The importance of the often-neglected lensing term, becomes more prominent at higher redshifts. At $z=1.5$ the error induced on $C_\ell^{\kappa g}$ grows to about  $(25-40)\%$ depending on the angular scale. For current measurements of $E_g$ which are performed at lower redshifts, for instance at $z=0.57$, the error induced by neglecting these extra contributions is small. However in order to take advantage of the high redshift information from future surveys such as Euclid to constrain $E_g$ with higher precision, we definitely have to include the lensing term. With this contribution the ratio $C^{\ka g}/(\beta C^{gg})$ is no longer scale invariant nor bias-independent since we have to add the bias independent $C^{\ka \ka}$ term to $C^{\ka g}$, which contributes a term inversely proportional to $b$  in $E_g$. Therefore, this quantity looses its  interest to a large extent. 

One way to circumvent this problem, might be to use the galaxy-galaxy lensing signal to subtract the $C^{\ka \ka}$ contribution.    Another very promising  possibility is to measure $E_g$ from intensity mapping, e.g., with 21-cm surveys and CMB lensing as proposed in~\cite{Pourtsidou:2015ksn}. 
The advantage of this is that because of flux conservations intensity maps have no lensing term at first order~\cite{Hall:2012wd}.
 
\section*{Acknowledgements}
We thank Tobias Baldauf for clarifying discussions, as well as Roy Maartens and Alkistis Pourtsidou  for pointing out relevant literature. A.M.D is supported by the Tomalla foundation and R.D. acknowledges support from the Swiss National Science Foundation.
We thank the Galileo Galilei Institute for Theoretical Physics for the hospitality and the INFN for partial support during the completion of this work.
\bibliography{bibEg}

\end{document}